\documentclass[preprint,aps,showpacs,showkeys]{revtex4}

\usepackage{graphics}
\usepackage{amsfonts}
\usepackage{amsmath}
\usepackage{epsfig}
\usepackage{epsf}
\usepackage{graphicx}
\usepackage{dcolumn}
\usepackage{bm}
\usepackage{color}

\begin{document}

\title{Nonperturbative numerical calculation of the fine and hyperfine structure of muonic hydrogen by Breit potential including the effects from the proton size}

\author{Hou-Rong Pang$^{1}$, Hai-Qing Zhou$^{1,2}$ \footnote{Email:zhouhq@seu.edu.cn}}

\affiliation{$^1$Department of Physics, Southeast University, Nanjing,210094\\
$^2$State Key Laboratory of Theoretical Physics, Institute of Theoretical Physics,\\
Chinese Academy of Sciences, Beijing\ 100190,\ P. R. China }


\begin{abstract}
By solving the two-body Schordinger equation in a very high precise nonperturbative numerical (NPnum) way,
we reexamine the contributions of fine, hyperfine structure splittings of muonic hydrogen based on the Breit potential. The comparison of our results with those by the first order perturbative theory ($^{1st}$PT) in the literature shows, when the structure of proton is considered, the differences between the results by the $^{1st}$PT and NPnum methods are small for the fine and hyperfine splitting of $2P$ state, while are about $-0.009$ meV and $0.08$ meV for the $F=1$ and total hyperfine splitting of $2S$ state of muonic hydrogen, respectively. These differences are larger than the current experimental precision and would be significant to be considered in the theoretical calculation.
\end{abstract}



\maketitle

\section{Introduction}


In 2010, a precision measurement \cite{nature466_213} of the Lamb shift
in muonic hydrogen
by using pulsed laser spectroscopy was performed and gave $E_{2P_{3/2}^{F=2}}^{Ex}-E_{2S_{1/2}^{F=1}}^{Ex}=206.2949$ meV. Combing this precise value
with the theoretical calculation \cite{nature466_213}
\begin{eqnarray}
E_{2P_{3/2}^{F=2}}^{Th}-E_{2S_{1/2}^{F=1}}^{Th} &=& 209.9779-5.2262r_p^2+0.0347r_p^3,
\label{equation:nature2010}
\end{eqnarray}
the values of the proton radius is extracted as $r_p=0.84184$ fm \cite{nature466_213}.
In 2013, the further precise measurements of $2S-2P$ transition frequencies of muonic hydrogen \cite{science339_417}
gave the magnetic radius of proton $r_M=0.87$ fm
and the charge radius $r_E=0.84087$ fm which are not significantly different from the value given by Ref. \cite{nature466_213}.

On the other hand, based on the hydrogen data or the $ep$ scattering data, CODATA-2010 gave $r_p \approx 0.878$ fm \cite{codata2010}, which is much larger than the results by the muonic hydrogen's Lamb-shift. And if this value of proton radius is used, the theoretical prediction for the Lambs shift of muonic hydrogen  gives \cite{physrep422_1}
\begin{eqnarray}
E_{2P_{3/2}^{F=2}}^{Th} -E_{2S_{1/2}^{F=1}} ^{Th}|_{r_p=0.878fm} = 205.9726~\textrm{meV},
\end{eqnarray}
which deviates from the experimental Lamb shift of muonic hypdrogen about $0.32$ meV.

Many theoretical calculations \cite{Boire2012}, data analysis \cite{Sick2012} and possible new mechanisms such as the three body physics \cite{three-body}, the new exotic particles interactions \cite{new-boson-exchange}, the higher-order contribution of the finite size \cite{high-order}\textit{ etc}., have been discussed to try to understand such discrepancy. And also new experiment of $ep$ scattering is proposed in JLab \cite{PRad-JLab}. Combining all these current analysis, briefly, the radius of proton is still not well understood.

For the muonic hydrogen, the energy  transition of $2P_{3/2}^{F=2}$ and $2S_{1/2}^{F=1}$ usually are expressed as
\begin{eqnarray}
E_{2P_{3/2}^{F=2}}^{Th}-E_{2S_{1/2}^{F=1}}^{Th} = \Delta E_{Lamb}^{2S-2P} +\Delta E_{FS}^{2P}+\frac{3}{8} \Delta E_{HFS}^{2P_{3/2}}-\frac{1}{4}\Delta E_{HFS}^{2S}.
\label{E2P2S}
\end{eqnarray}
In the literature, the contributions of the four terms are usually calculated by the perturbative theory. Using the quasipotential method in quantum electrodynamics \cite{quasipotential}, the contributions to the four terms can be expressed as \cite{nature466_213,annals326_500,Martynenko2005,Martynenko2008}
\begin{figure}[htbp]
\center{\epsfxsize 5.0 truein\epsfbox{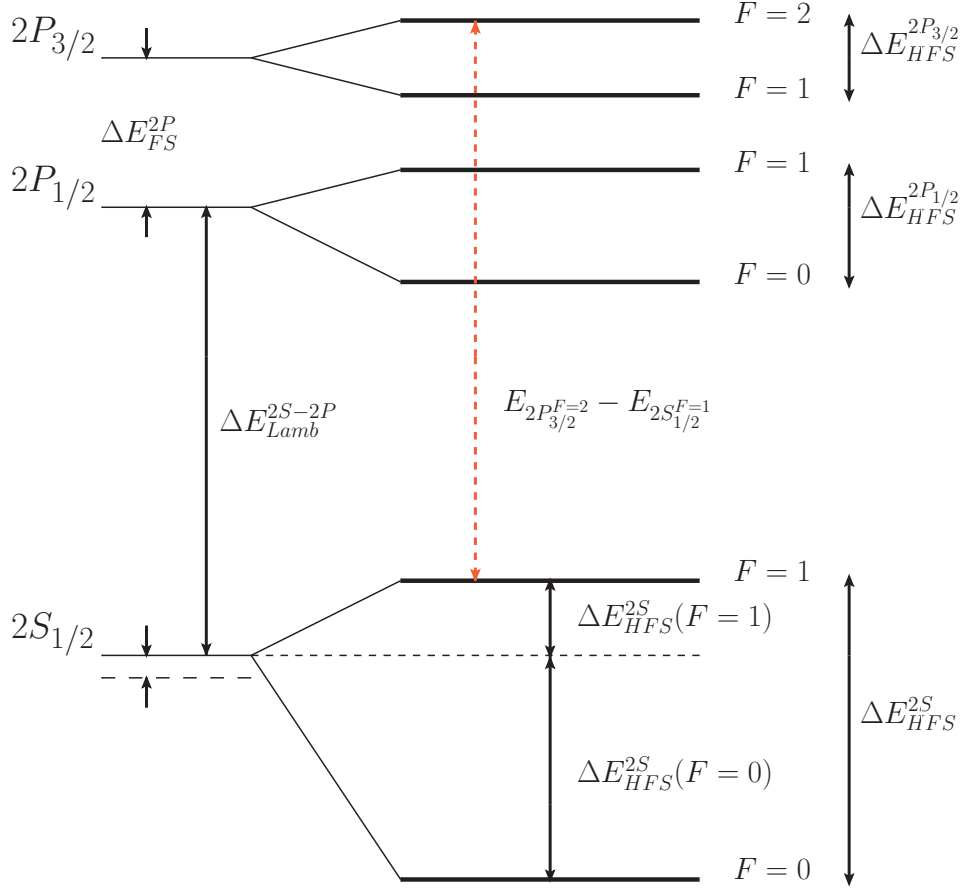}}
\caption{ The 2S and 2P energy levels of muonic hydrogen.}
\label{figure:1}
\end{figure}

\begin{eqnarray}
\Delta E_{Lamb}^{2S-2P} &=& \Delta E_{ovp} + \Delta E_{KS} + \Delta E_{SE} +\Delta E_{QCD} -5.2262r_p^2 +0.0347r_p^3 \nonumber \\
\Delta E_{HFS}^{2S} &=&\Delta E^{2S,B}_{HFS}+\Delta E^{2S,AMM}_{HFS}+\Delta E^{2S,other}_{HFS} \nonumber \\
\Delta E_{HFS}^{2P_{3/2}} &=& \Delta E^{2P_{3/2},B}_{HFS}+\Delta E^{2P_{3/2},AMM}_{HFS}+\Delta E^{2P_{3/2},other}_{HFS} \nonumber \\
\Delta E_{FS}^{2P} &=& \Delta E^{2P,B}_{FS}+\Delta E^{2P,AMM}_{FS}+\Delta E^{2P,other}_{FS}
\end{eqnarray}
where $\Delta E_{ovp}$ ($205.0074$ meV by the first order  perturbative theory ($^{1st}$PT) and $0.1509$ meV by the second order perturbative theory), $\Delta E_{KS}$ ($1.5081$ meV) and $\Delta E_{SE}$ ($-0.6677$ meV) are the energy shifts due to the one-loop vacuum polarization, two-loop vacuum polarization and the sum of self-energy and muonic-vacuum polarization, correspondingly, $\Delta E_{QED}$ ($0.0586$ meV) is the energy shift due to all further QED corrections,
and the last two terms in $\Delta E_{Lamb}^{2S-2P}$ are relevant radius-dependent contributions\cite{annals326_500}, $\Delta E^{2S,B}_{HFS}$ ($22.8054$ meV), $\Delta E^{2P_{3/2},B}_{HFS}$ ($3.392112$ meV) and $\Delta E^{2P,B}_{FS}$ ($8.329150$ meV) are the Fermi energies, $\Delta E^{2S,AMM}_{HFS}$ ($0.0266$ meV), $\Delta E^{2P_{3/2},AMM}_{HFS}$ ($-0.000886$ meV)  and $\Delta E^{2P,AMM}_{FS}$ ($0.017637$ meV) are the contributions from the anomalous magnetic moment of muon, $\Delta E^{2S,other}_{HFS}$ , $\Delta E^{2P_{3/2},other}_{HFS}$ and $\Delta E^{2P_{3/2},other}_{FS}$ are the other contributions\cite{Martynenko2005,Martynenko2008}.

Some of the above perturbative results have been checked by the nonperturbative numerical (NPnum) calculations, for example within the framework of the multiconfiguration Dirac-Fock (MCDF) method in \cite{Indelicato2013} and shotting-like method using quad-precision Fortran in \cite{Carroll2011}. In this work, by using the Mathematica, we present another high precise NPnum calculations (much more precise than the quad-precision) on the energy shifts
$E^{2S,B+AMM}_{HFS}$, $E^{2P_{3/2},B+AMM}_{HFS}$ and $E^{2P,B+AMM}_{FS}$ with considering the effects from the proton structure.
And as a comparison, also the calculation of $\delta E_{ovp}$ is presented.

\section{Formula and numerical method}

The leading order contribution to the fine and hyperfine structure of muonic hydrogen due to the proton structure is from the two photon exchange diagrams. Since there is IR divergence in these diagrams and such IR divergence is not dependent on the proton structure (only dependent on its charge), the one photon exchange diagram should be considered together in some way to cancel such IR divergence. This leads to the complexity in the discussion of the numerical calculation, so at present we take the effective potential from the one photon exchange diagram as an example to discuss the difference between the $^{1st}$PT calculation and the NPnum calculation.
\begin{figure}[htbp]
\center{\epsfxsize 3.0 truein\epsfbox{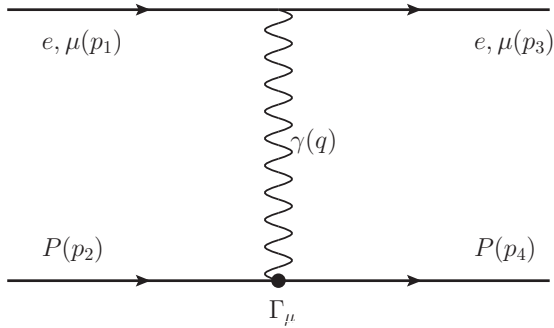}}
\caption{ One photon exchange Feynman diagram considering the form factors of proton.}
\label{figure:one-photon-exchange}
\end{figure}

The Feynman diagram for one photon exchange with the proton structure is showed as Fig. \ref{figure:one-photon-exchange}, where the vertex of $\gamma NN$ is taken as
\begin{eqnarray}
\Gamma_{\mu}&=&ie(F_1\gamma_{\mu}+\frac{iF_2}{2m_p}\sigma_{\mu\nu}q^{\nu}),
\end{eqnarray}
with  $F_1=(G_E+\tau G_M)/(1+\tau)$,$F_2=(G_M-G_E)/(1-\tau)$, $\tau=Q^2/4m_p^2$, $G_{E,M}$ the electromagnetic form factors of proton, $m_p$ the mass of proton, and $q=p_3-p_1$ the four momentum of exchanged photon. The correction to the Coulomb potential from this diagram was discussed in the recent work \cite{Pascalutsa2015}, and in this work, we discuss its corrections to the fine and hypefine Breit potential by a precise numerical method. By using the quasipotential method the Breit potential can be expressed as \cite{annals326_500,Kelkar2012}.

\begin{eqnarray}
V_{ovp}(r)&=&-\frac{\alpha^2}{\pi r} \int_1^{\infty} du e^{-2um_e r}\frac{ \sqrt{ (u^2-1)}(2u^2+1)}{3u^4}, \nonumber \\
V_{fs}(r)&=&\frac{\alpha}{2m_{\mu}^2}[(1+2\kappa_{\mu})+\frac{2m_{\mu}}{m_p}(1+\kappa_{\mu})](\frac{1}{r^3}+\frac{G_{fs}}{r^3}){\bf L} \cdot {\bf S_{\mu}}, \nonumber \\
V_{HFS}^{S wave}(r)&=&\frac{\alpha \mu_p \mu_{\mu}}{4r^3m_{\mu} m_p}{\bm \sigma}_{\mu} \cdot {\bm \sigma}_p \frac{m^3 r^3}{3} e^{-mr}, \nonumber \\
V_{HFS}^{P wave}(r)&=&\frac{\alpha \mu_p \mu_{\mu}}{4r^3m_{\mu} m_p}[(3{\bm \sigma}_{\mu} \cdot {\bm {\hat r}}{\bm \sigma}_p \cdot {\bm{\hat r}}(1+h_1)-{\bm \sigma}_{\mu} \cdot {\bm \sigma}_p(1+h_2)) \nonumber \\
&& + 2{\bf L} \cdot {\bm \sigma}_p(\frac{1+h_3}{\mu_{\mu}}+\frac{m_{\mu}}{2m_p}\frac{h_4}{\mu_p \mu_{\mu}})],
\end{eqnarray}
where the One-loop Uehling potential $V_{ovp}$ is also presented for comparison and
\begin{eqnarray}
G_{fs}&=&-(1+\frac{\kappa_p}{(1-k^2)^2})e^{-mr}(1+mr)-(1+\frac{\kappa_p}{1-k^2})\frac{m^2}{2}r^2e^{-mr} \nonumber\\
&&+\frac{\kappa_p}{(1-k^2)^2}e^{-mkr}(1+mkr),\nonumber
\end{eqnarray}
\begin{eqnarray}
h_1=-e^{-mr}(1+mr)-\frac{m^2r^2}{2}e^{-mr}-\frac{m^3r^3}{6}e^{-mr},\nonumber
\end{eqnarray}
\begin{eqnarray}
h_2=-e^{-mr}(1+mr)-\frac{m^2r^2}{2}e^{-mr}-\frac{m^3r^3}{2}e^{-mr},\nonumber
\end{eqnarray}
\begin{eqnarray}
h_3=-e^{-mr}(1+mr)-\frac{m^2r^2}{2}e^{-mr},\nonumber
\end{eqnarray}
\begin{eqnarray}
h_4&=&(1+2\kappa_p)(1+h_3)+\frac{\kappa_p}{(1-k^2)^2}e^{-mr}(1+mr)+\nonumber \\ &&\frac{\kappa_p}{1-k^2}\frac{m^2r^2}{2}e^{-mr}-\frac{\kappa_p}{(1-k^2)^2}e^{-mkr}(1+mkr),\nonumber
\end{eqnarray}
with $\alpha$ the fine-structure constant, $\mu_p$ and $\mu_\mu$ the anomalous magnetic moments of proton and muon, $m$ the parameter in the electromagnetic form factors of proton which can be related with proton size as $r_p^2=12/m^2$, $k=2m_p/m$, $m_e$, $m_\mu$ the mass of electron and muon. To include the effects from the proton structure in the above Breit potential, the electromagnetic form factors of proton $G_{E,M}(Q^2)$ have been taken approximately as the usual dipole form as \cite{Kelkar2012}
\begin{eqnarray}
G_E(Q^2)=G_M(Q^2)/\mu_p = 1/(1+Q^2/m^2)^2,
\end{eqnarray}
with $Q^2=-q^2$. The contributions of spin related operators in the Breit potential to the $2S$ and $2P$ states are listed in Tab. \ref{table:spin-contribution}.

\begin{table}
\begin{tabular}{|p{2cm}<{\centering}||p{3cm}<{\centering}|p{3cm}<{\centering}|p{3cm}<{\centering}|}\hline
&$3({\bm \sigma}_\mu \cdot {\hat {\bf r}})({\bm \sigma}_p \cdot
{\hat {\bf r}})$&${\bm \sigma}_p \cdot {\bm \sigma}_\mu$&$2{\bf L} \cdot {{\bm \sigma}_p}$  \\
\hline
$2S_{1/2}^{F=0}$&$-3$ & $-3$&$0$\\ \hline
$2S_{1/2}^{F=1}$&$1$ & $1$&$0$\\ \hline
$2P_{1/2}^{F=0}$&$-3$ & $1$&$-4$\\ \hline
$2P_{1/2}^{F=1}$&$1$ & $-\frac{1}{3}$&$\frac{4}{3}$\\ \hline
$2P_{3/2}^{F=2}$&$\frac{3}{5}$ & $1$&$2$\\ \hline
$2P_{3/2}^{F=1}$&$-1$ & $-\frac{5}{3}$&$-\frac{10}{3}$\\ \hline
\end{tabular}
\caption{\label{table:spin-contribution} The contributions of the spin related operators to $2S$ and $2P$ states.}
\end{table}
When the effects from the proton structure are neglected in the above effective potentials (taking $m \rightarrow \infty$), the energy shifts by $^{1st}$PT reproduce the same $E^{2S,B+AMM}_{HFS}$, $E^{2P_{3/2},B+AMM}_{HFS}$ and $E^{2P,B+AMM}_{FS}$ with those used in the literatures\cite{Martynenko2005,Martynenko2008}. The corrections from the proton structure in the above Breit potential is corresponding to replace the zero momentum transfer approximation by including the full $q^2$ dependence of the electromagnetic form factors of proton in the one photon exchange Feynman diagram. This is different with the corrections from the two photon exchange diagrams discussed in \cite{Martynenko2005,Martynenko2008}. Since in this work our focus is on the difference between the results by $^{1st}$PT and NPnum calculation, we take the Breit potential as an example to show the difference.

In our numerical calculation, we use the shotting method to find out the energy spectrum by solving the reduced Schrodinger equations for $u(r)$ directly, with $u(r)=R(r)r$, $\psi({\bf r})=R(r)Y(\theta,\phi)$ and $\psi({\bf r})$ the wave function. In the detail, for muonic hydrogen we take approximately $u(r=0)\approx u(r=10^{-50}fm)$ and $u(r=\infty) \approx u(r=10^{6}{fm})$ to simulate the behaviors of wave function at the boundary. We keep 200 digits of the numbers in the calculation and take the PrecisionGoal of NDSolve in Mathematica as 15, respectively. By these approximations, as a check we reproduce the energy spectrum of muonic hydrogen under the Coulomb potential with the precision better than $10^{-10}$ meV. Since the numerical calculation is based on the shotting method, the precision is not sensitive on the form of the potentials or the solutions, this is different with the basis expansion method or variational methods usually used, and ensures our numerical calculation reliable for the other potentials.

\section{numerical result}
In Tab. \ref{table:energy-shift-2P} and \ref{table:energy-shift-2S}, we present the results by the $^{1st}$PT and NPnum calculation including the effects from the proton structure in the Breit potential. The corresponding results without considering the proton structure by $^{1st}$PT and from the one-loop Uehling potential are also presented for comparison.

\begin{table}[htbp]
\begin{tabular}
{|p{1.5cm}<{\centering}|p{1.5cm}<{\centering}|p{1.5cm}<{\centering}|p{1.5cm}<{\centering}|p{1.5cm}<{\centering}|p{1.7cm}<{\centering}|p{1.7cm}<{\centering}|}
\hline
\raisebox{-1.50ex}[0cm][0cm]{$r_p$(fm)}&
\multicolumn{2}{p{3.0cm}<{\centering}|}{$\Delta E_{FS}^{2P}$} &
\multicolumn{2}{p{3.0cm}<{\centering}|}{$\Delta E_{HFS}^{2P^{3/2}}$} &
\multicolumn{2}{p{3.0cm}<{\centering}|}{$\Delta E_{ovp}$}  \\
\cline{2-7}
&$^{1st}$PT &NPnum  &$^{1st}$PT   &NPnum  &$^{1st}$PT   &NPnum \\
\hline
0&8.34676 &-& 3.39121  &- &205.00659&205.15747\\ \hline
0.83112&8.34673&8.34703&3.39121 &3.39123 &205.00659&205.15747\\ \hline
0.84184&8.34672&8.34702 &3.39121 &3.39123 &205.00659&205.15747\\ \hline
0.87800&8.34672&8.34702 &3.39121 &3.39123  &205.00659&205.15747\\ \hline
\end{tabular}
\caption{\label{table:energy-shift-2P} Energy shifts of different potentials using perturbative and precise numerical calculations in meV where $^{1st}$PT and NPnum denote the first order pertubative and precise nonperturbative numerical calculation, respectively. The typical proton size $r_p=0,0.83112,0.84184,0.878 fm$ are taken as examples for comparison. The results by $^{1st}$PT are same with those in \cite{annals326_500,Martynenko2005}.}
\end{table}

\begin{table}[htbp]
\begin{tabular}
{|p{1.5cm}<{\centering}|p{1.5cm}<{\centering}|p{1.5cm}<{\centering}|p{1.7cm}<{\centering}|p{1.7cm}<{\centering}|p{1.5cm}<{\centering}|p{1.5cm}<{\centering}|p{1.7cm}<{\centering}|p{1.7cm}<{\centering}|}
\hline
\raisebox{-1.50ex}[0cm][0cm]{$r_p$(fm)}&
\multicolumn{2}{p{3.0cm}<{\centering}|}{$\Delta E_{HFS}^{2S}(F=1)$} &
\multicolumn{2}{p{3.0cm}<{\centering}|}{$\Delta E_{HFS}^{2S}(F=0)$} &
\multicolumn{2}{p{3.0cm}<{\centering}|}{$\frac{\Delta E_{HFS}^{2S}(F=0)}{\Delta E_{HFS}^{2S}(F=1)}$} & \multicolumn{2}{p{3.0cm}<{\centering}|}{$\Delta E_{HFS}^{2S}$}\\
\cline{2-9}
&$^{1st}$PT &NPnum & $^{1st}$PT &NPnum &$^{1st}$PT &NPnum &$^{1st}$PT &NPnum \\
\hline
0 & 5.70798  & - & 17.12394 &- &3&- &22.83192&-\\ \hline
0.83112 & 5.67921 & 5.66943 &17.03762 &17.12628 &3&3.0208 &22.71683&22.79571\\ \hline
0.84184 & 5.67884 & 5.66918 &17.03651 &17.12406 &3&3.0206 &22.71535&22.79324\\ \hline
0.87800 & 5.67759 & 5.66832 &17.03277 &17.11676 &3& 3.0197 &22.71036&22.78508\\ \hline
\end{tabular}
\caption{\label{table:energy-shift-2S} Energy shifts of different potentials using perturbative and precise numerical calculations in meV. The notations are same with Tab. \ref{table:energy-shift-2P}. The results by $^{1st}$PT are same with those in \cite{Martynenko2008}.}
\end{table}

From the last two columns of Tab. \ref{table:energy-shift-2P}, we see the results by $^{1st}$PT and NPnum calculations give about $0.15088$ meV difference for $V_{ovp}$. Actually the second order perturbative calculation of $V_{ovp}$ gives the contribution about $0.1509$ meV \cite{annals326_500}, so the combination of the first and second order pertubative calculation of $V_{ovp}$  is almost same with our NPnum calculation. When including the effects from the proton structure and taking $r_p=0.83112$ fm, $0.84184$ fm, $0.878$ fm as examples, the differences between the $^{1st}$PT and precise NPnum calculations are about $3\times10^{-4}$ and $2\times10^{-5}$ meV for the fine and hyperfine splitting of $2P$ states, respectively, and also these two splittings are almost independent on the proton size in the region $r_p\in [8.2,9.0]$ fm. From the Tab. \ref{table:energy-shift-2S}, we see the derivations of the $^{1st}$PT and NPnum calculation are about $-0.009$ and $0.02$ meV for the hyperfine splitting of $2S$ state $\Delta E_{HFS}^{2S}(F=1)$ and $\frac{1}{4}\Delta E_{HFS}^{2S}$, respectively, which should not be omitted comparing with the precision of current experiments. We want to emphasize that by the NPnum calculation, the ratios $\frac{\Delta E_{HFS}^{2S}(F=0)}{\Delta E_{HFS}^{2S}(F=1)}$ are not strictly equal to $3$ as predicted by the $^{1st}$PT, but are about $3.02$ as showed in Tab. \ref{table:energy-shift-2S} and the absolute difference between $\Delta E_{HFS}^{2S}(F=1)$ and $\frac{1}{4}\Delta E_{HFS}^{2S}$ are large. Such a discrepancy means the relation $\Delta E_{HFS}^{2S}(F=1)=\frac{1}{4}\Delta E_{HFS}^{2S}$ is not suitable to be used as usual. To show the results in a more direct way, we use the polynome of $r_p$ to fit the numerical results of $\Delta E_{HFS}^{2S}(F=1)$ in the region $r_p\in [8.2,9.0]$ fm by taking one point every $0.001$ fm and the results are expressed as
\begin{eqnarray}
\Delta E_{HFS}^{2S}(F=1,\textrm{NPnum}) &=& 5.68008-0.00261343r_p-0.0122699r_p^2~~~~~~~\textrm{meV} \nonumber\\
\Delta E_{HFS}^{2S}(F=1,^{1st}\textrm{PT}) &=& 5.70793-0.0346149r_p+0.0000613066r_p~~~~\textrm{meV}
\end{eqnarray}
with the residual mean square of the fitting as small as about $10^{-11}$ meV$^2$.

After including the difference of $\Delta E_{HFS}^{2S}(F=1)$ by our estimation, the theoretical energy shift Eq.(\ref{equation:nature2010}) is changed as
\begin{eqnarray}
E_{2P_{3/2}^{F=2}}^{Th}-E_{2S_{1/2}^{F=1}}^{Th} &=& 209.9869-5.2262r_p^2+0.0347r_p^3
\end{eqnarray}
and if $r_p$ is taken as $0.878$fm, then the modified $E_{2P_{3/2}^{F=2}}^{Th}-E_{2S_{1/2}^{F=1}}^{Th}$ is estimated as $205.9816$ meV, which deviates from the experimental results about $0.31$ meV. This means after using the precise NPnum calculation of $\Delta E_{HFS}^{2S}(F=1)$, the discrepancy of the measurement of the proton size is reduced about 3\%, while if we take the $\frac{1}{4}\Delta E_{HFS}^{2S}$ as input then the energy shift Eq.(\ref{equation:nature2010}) is changed as
\begin{eqnarray}
E_{2P_{3/2}^{F=2}}^{Th}-E_{2S_{1/2}^{F=1}}^{Th} &=& 209.9579-5.2262r_p^2+0.0347r_p^3
\end{eqnarray}
and if $r_p$ is taken as $0.878$fm, then $E_{2P_{3/2}^{F=2}}^{Th}-E_{2S_{1/2}^{F=1}}^{Th}$ is estimated as $205.9526$ meV, which deviates from the experimental results about $0.34$ meV. We see the discrepancy will be intensified about 7\%. The full results show we should be careful to deal with the splitting beyond the $^{1st}$PT and should replace $\frac{1}{4}\Delta E_{HFS}^{2S}$ by $\Delta E_{HFS}^{2S}(F=1)$ in the calculation.

According to Tab. \ref{table:spin-contribution} and Fig. \ref{figure:1}, we also have
\begin{eqnarray}
E_{2P_{3/2}^{F=1}}-E_{2S_{1/2}^{F=0}}=\Delta E_{Lamb}^{2S-2P}-\frac{5}{8}\Delta E_{HFS}^{2P_{3/2}}+\Delta E_{FS}^{2P}+\frac{3}{4}\Delta E_{HFS}^{2S},
\end{eqnarray}
and
\begin{eqnarray}
\Delta E=(E_{2P_{3/2}^{F=1}}-E_{2S_{1/2}^{F=0}})-(E_{2P_{3/2}^{F=2}}-E_{2S_{1/2}^{F=1}})=\Delta E_{HFS}^{2S}-\Delta E_{HFS}^{2P_{3/2}}.
\end{eqnarray}
We note that $\Delta E$ depends on $\Delta E_{HFS}^{2S}$  instead of $\Delta E_{HFS}^{2S}(F=1)$ appeared in $E_{2P_{3/2}^{F=2}}-E_{2S_{1/2}^{F=1}}$, so for $\Delta E$ the NPnum calculation gives relative larger corrections. We present the results of $\Delta E$ from the Breit potential by the two methods in Tab. \ref{table:DeltaE}. With the different proton radius as input, the differences of $\Delta E$ are about $10^{-3}\sim10^{-2}$ meV. At present, the $E_{2P_{3/2}^{F=2}}-E_{2S_{1/2}^{F=1}}$ and $E_{2P_{3/2}^{F=1}}-E_{2S_{1/2}^{F=0}}$ transition frequency in muonic hydrogen have been both measured with very high accuracy \cite{science339_417}, the experimental value of $\Delta E$ is about $19.56$ meV. Our numerical results are different from the experimental value since we have not considered the other corrections beside Breit potential. However, our calculation implies the precise NPnum calculation is needed.

\begin{table}[htbp]
\begin{tabular}
{|p{2cm}<{\centering}|p{2cm}<{\centering}|p{2cm}<{\centering}|}
\hline
\raisebox{-1.50ex}{$r_p$(fm)}&
\multicolumn{2}{|c|}{$\Delta E(\textrm{meV})$}  \\
\cline{2-3} &$^{1st}$PT  &NPnum  \\ \hline
0&$19.44070$&- \\ \hline
0.83112&$19.32562$&$19.40434$ \\ \hline
0.84184&$19.32414$&$19.40187$ \\ \hline
0.87800&$19.31915$&$19.39371$ \\ \hline
\end{tabular}
\caption{\label{table:DeltaE} The difference of $2S-2P$ transition energy  $\Delta E$ of muonic hydrogen in meV with different radius of proton as input. The notations are same with Tab. \ref{table:energy-shift-2P}.}
\end{table}

Since the obvious difference exists between the $^{1st}$PT and precise NPnum calculations of the hyperfine splitting of muonic hydrogen's $2S$ state, we also present the similar comparison of the $ep$ system in Tab. \ref{table:HFS-hydrogen} where the results show the calculation for the hyperfine splitting of $S$ state  in $ep$ system by the $^{1st}$PT  is also not good enough. Different with $^{1st}$PT, the results by the precise NPnum method are more sensitive on the proton size.

\begin{table}
\begin{tabular}{|p{2cm}<{\centering}||p{2cm}<{\centering}|p{2cm}<{\centering}||p{2cm}<{\centering}|p{2cm}<{\centering}|}
\hline
\raisebox{-1.50ex}[0cm][0cm]{$r_p$(fm)}&
\multicolumn{2}{c||}{$E_{HFS}^{1S}(kHz)$} &
\multicolumn{2}{c|}{$E_{HFS}^{2S}(kHz)$}  \\
\cline{2-5}& $^{1st}$PT &NPnum &$^{1st}$PT &NPnum \\
\hline
0       &$1420478.8$& -&$177559.8 $& - \\
\hline
0.83112 &$1420440.1$  &$1425769.6$ &$177555.0 $  &$178221.2$ \\
\hline
0.84184 &$1420439.6$  &$1425700.8$ &$177555.0 $  &$178212.6$ \\
\hline
0.87800 &$1420438.0$  &$1425481.1$  &$177554.7$  &$178185.1$ \\
\hline
\end{tabular}
\caption{\label{table:HFS-hydrogen}Frequencies in kHz of hyperfine splitting $E_{HFS}^{1S}$ and $E_{HFS}^{2S}$ in hydrogen. }
\end{table}

In a summary, by using shotting method in Mathematica we give a very high precise NPnum calculation for the energy shifts of the
Breit potential including the effects from the proton structure.
Our results show that when taking into account the proton structure,
the precise NPnum calculations give very small corrections to the hyperfine splitting of $2P_{3/2}$ and fine structure of $2P$ states, but give about $-0.009$ meV and $0.08$ meV differences for the hyperfine splitting $\Delta E_{HFS}^{2S}(F=1)$ and $\Delta E_{HFS}^{2S}$ of munoic hydrogen with those usually used in the literatures
by $^{1st}$PT. The similar properties are also found in the hydrogen case.

\section{Acknowledgments}
This work is supported by the National Sciences Foundations of China
under Grant No. 11375044 and in part by the Fundamental Research Funds for the Central Universities under Grant No. 2242014R30012.


\end{document}